\documentclass[conference]{IEEEtran}
\IEEEoverridecommandlockouts
\usepackage{cite}
\usepackage{amsmath,amssymb,amsfonts}
\usepackage{algorithmic}
\usepackage{graphicx}
\usepackage{textcomp}
\usepackage{xcolor}
\usepackage{epstopdf}
\usepackage{multirow}
\usepackage{subfigure}
\usepackage{url}
\usepackage[ruled,vlined]{algorithm2e}
\include{pythonlisting}
\def\BibTeX{{\rm B\kern-.05em{\sc i\kern-.025em b}\kern-.08em
    T\kern-.1667em\lower.7ex\hbox{E}\kern-.125emX}}

\begin{document}
\title{Towards Learning-automation IoT Attack Detection through Reinforcement Learning
}

\author{
    \IEEEauthorblockN{
        Tianbo Gu\IEEEauthorrefmark{1},
        Allaukik Abhishek\IEEEauthorrefmark{2},
        Hao Fu\IEEEauthorrefmark{1}, 
        Huanle Zhang\IEEEauthorrefmark{1},
        Debraj Basu \IEEEauthorrefmark{1},
        Prasant Mohapatra\IEEEauthorrefmark{1}
    }
    \IEEEauthorblockA{
        \IEEEauthorrefmark{1}
        Department of Computer Science, University of California, Davis, CA, USA\\
        \IEEEauthorrefmark{2}
        ARM Research, Austin, TX, USA\\
        Email: \{tbgu, haofu, dtczhang, dbasu, pmohapatra\}@ucdavis.edu, Allaukik.Abhishek@arm.com
    }
}

% \IEEEoverridecommandlockouts
% \IEEEpubid{\makebox[\columnwidth]{978-1-7281-7374-0/20/\$31.00~\copyright2020 IEEE \hfill} \hspace{\columnsep}\makebox[\columnwidth]{ }}

%978-1-7281-7374-0/20/$31.00 ©2020 IEEE

\maketitle
%\IEEEpubidadjcol
\begin{abstract}

As a massive number of the Internet of Things (IoT) devices are deployed, the security and privacy issues in IoT arouse more and more attention. The IoT attacks are causing tremendous loss to the IoT networks and even threatening human safety. Compared to traditional networks, IoT networks have unique characteristics, which make the attack detection more challenging. First, the heterogeneity of platforms, protocols, software, and hardware exposes various vulnerabilities. Second, in addition to the traditional high-rate attacks, the low-rate attacks are also extensively used by IoT attackers to obfuscate the legitimate and malicious traffic. These low-rate attacks are challenging to detect and can persist in the networks. Last, the attackers are evolving to be more intelligent and can dynamically change their attack strategies based on the environment feedback to avoid being detected, making it more challenging for the defender to discover a consistent pattern to identify the attack.

In order to adapt to the new characteristics in IoT attacks, we propose a reinforcement learning-based attack detection model that can automatically learn and recognize the transformation of the attack pattern. Therefore, we can continuously detect IoT attacks with less human intervention. In this paper, we explore the crucial features of IoT traffics and utilize the entropy-based metrics to detect both the high-rate and low-rate IoT attacks. Afterward, we leverage the reinforcement learning technique to continuously adjust the attack detection threshold based on the detection feedback, which optimizes the detection and the false alarm rate. We conduct extensive experiments over a real IoT attack dataset and demonstrate the effectiveness of our IoT attack detection framework.

\end{abstract}

\begin{IEEEkeywords}
Internet of Things, Intrusion detection, Reinforcement learning, Anomaly detection, IoT security, Entropy, Wireless traffic, Artificial Intelligence.
\end{IEEEkeywords}
\section{Introduction}

Based on the report from \cite{iotnum}, there will be 75 billion Internet of Things (IoT) connected devices in the world by 2025. The growth of IoT devices has exploded over the past ten years. IoT techniques have been applied to various fields, such as smart home, smart city, and smart industry. The IoT networks dramatically enhance the productivity, quality, and efficiency of work and create substantial economic profits. For instance, an IoT application can automatically open the home window or the air conditioner once the indoor temperature exceeds a threshold, making our lives more automatic without human intervention. 

The immense network comprising of IoT devices is interrelated and independent with the current Internet. However, the corresponding security and privacy solutions do not keep up with the increasing scale of IoT devices. The vast commercial market also earns the attention of attackers. The attackers attempt to explore and launch a variety of attacks in IoT networks, which may cause enormous economic loss and pose a severe threat to security and privacy. However, the unique characteristics in IoT networks make it more challenging to provide comprehensive security and privacy solutions.

First, to maintain a long life cycle and combat the limitation of computations and battery capacity, most IoT devices have to be compromised with security and privacy issues. Some IoT devices adopt weak encryption or even no encryption process. Meanwhile, due to no consistent standard, there are many IoT platforms and communication protocol proposed in the IoT network. The common IoT platforms include Samsung SmartThings\cite{smart2015}, Google Home\cite{google2015}, Apple HomeKit\cite{apple2015}, etc. These platforms also use different wireless communication protocols, such as \textit{WiFi, ZigBee, Z-Wave, BLE, etc.} However, the heterogeneity of platforms and communication protocols exposes various attack interfaces, thereby making it considerably challenging for a defender to safeguard against the attacks \cite{granjal2015security}.

Second, most IoT devices transmit the data at a relatively low rate. In order to avoid being detected, the attackers launch the intrusion with a low-rate trait and produce malicious traffic that blends in the normal traffic, which is challenging to be detected with superior disguise. Furthermore, with the rapid development of \textit{Artificial Intelligence} and \textit{Machine Learning}, the attacker is gradually possessing the capabilities of learning-automation and becoming smarter. To delay the detection time, the attacker can study the feedback from the environment and dynamically alter their attack vectors or launch a new evolved attack to avoid intrusion detection. The detected attack pattern by the defender will be invalid immediately, and the defender has to make an instantaneous response and recognize the new attack pattern fast.

Currently, most existing detection approaches leverage the traditional anomaly detection model \cite{garcia2009anomaly} without considering the unique characteristics in the IoT domains. Most detection methods do not consider the diversity of attack interfaces and can only detect very few attack types. Some researchers use deep learning \cite{7946998, mirsky2018kitsune, chaabouni2019network} to implement the network intrusion detection. But the computation power and time cost of these approaches are too expensive to be executed on a simple IoT gateway. Also, the existing detection approaches are not sensitive to low-rate attacks. Moreover, these solutions cannot adapt to the learning-automation attacks in IoT \cite{ll2016distributed, caminero2019adversarial}. The expensive detection process of attack patterns may be immediately invalid if the attackers continuously change their intrusion strategies. Therefore, we propose a generalized solution that can detect a variety of attacks by analyzing the features of IoT traffic. A lightweight detection model is designed using an entropy-based detection approach, which is more sensitive to low-rate attacks and has a low computation cost. More importantly, we propose a reinforcement learning-based IoT attack detection framework that can detect the evolving IoT attacks with more effectiveness. 

\textbf{Contribution:} In summary, our contributions break down into the following aspects:
\begin{itemize}
    \item We thoroughly explore all the unique attack characteristics in IoT networks, such as the low-rate attack, and propose a lightweight entropy-based attack detection approach that can efficiently detect various attacks.
    \item To resist the more intelligence of IoT attacks, we propose a reinforcement learning-based attack detection framework in IoT networks, which can automatically adjust the attack detection threshold and quickly discover the attacks even with the continuous transformation and evolution of IoT attacks.
    \item To the best of our knowledge, we are the first to incorporate reinforcement learning to develop a novel attack detection framework in IoT networks. Our approach can provide smarter defenses with less human supervision, which can greatly facilitate the current research in IoT security.
    \item We design and conduct extensive experiments on a $\sim 60$ GB IoT dataset involving a dozen IoT devices in the market and various types of IoT attacks. The evaluation demonstrates that our proposed IoT attack detection framework can adapt to the new attack characteristics and exhibit significant improvement in system utilities.
\end{itemize}

\textbf{Road map:} The rest of the paper is organized as follows. In Section \ref{sec:adversary}, we discuss the adversary model for IoT attack. Section \ref{sec:model} describes the IoT attack detection framework including the model and algorithm. We present our evaluation in Section \ref{sec:eval}. Related work and conclusion are presented in Section \ref{sec:relatedwork} and \ref{sec:conclusion} respectively.

\section{Adversary model}
\label{sec:adversary}

In this work, a typical IoT network (such as smart home, smart industry, smart city) is considered, where the IoT devices are first connected to the IoT gateway and then connected to the Internet, thereby forming a vast IoT network. The adversary scans the IoT devices and networks and explores the existing vulnerability that IoT networks have exposed. The exposed attack interfaces could include the bugs in IoT platform, the weak encryption vulnerabilities in wired and wireless communication protocols, the defects of IoT devices' hardware and software, etc.

We divide the attack types into two categories: (1) direct attacks and (2) reflection attacks. The direct attacks to IoT include TCP/UDP flooding, ARP spoofing, Ping of Death, etc. The reflection attacks include TCP SYN, Smurf, SNMP, etc. Some attacks attempt to consume all the available resources in IoT devices by attacking the network protocols in different layers, rendering the services of the IoT system unavailable. Some other attacks can exploit IoT networks' security vulnerabilities and try to compromise other devices in the network or inject false information in the system.

Moreover, most attacks need to persist in the IoT network for an extended period in order to operate effectively, such as DDoS attacks. To persist in the network, the attacks launched by the adversary usually have two properties: \textit{stealth} and \textit{resilience}. The adversary produces malicious traffic with no difference with legitimate traffic, thereby remaining elusive, such as low-rate attacks. On the other hand, some attackers become more intelligent and can transform or evolve their attack procedures via observing the environment's feedback, which could make it more challenging to identify the IoT attacks and sustain the resilience.

Our goal is to design an attack detection framework that considers the unique attack characteristics in IoT. The detection framework can identify IoT attacks with high superiority. For instance, the low-rate attacks sneak their malicious traffic into normal traffic, thereby avoiding detection. Furthermore, the defense strategies need to be upgraded and evolved as the development of learning-automation for attackers. The adversary could transform their attack strategies or develop a new attack approach based on the environment's feedback. The more intelligent and flexible attacks significantly increase the difficulties of defense. An efficient defensive scheme to mitigate attacks and secure IoT systems should be designed and implemented with considering the new attack characteristics in IoT.

\begin{figure*}
    \begin{minipage}[c]{0.6\linewidth}
    \includegraphics[height=33mm]{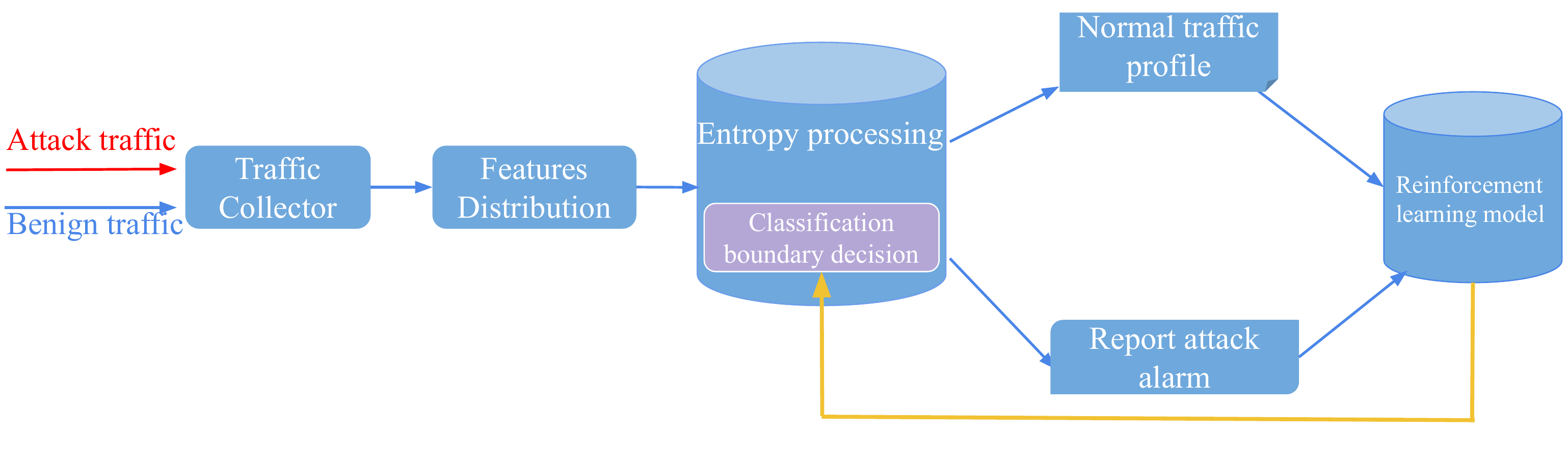}
    \caption{Framework of IoT attack detection via reinforcement learning.}
    \label{fig:framework}
    \end{minipage}
    \hfill
    \begin{minipage}[c]{0.35\linewidth}
    \includegraphics[height=33mm]{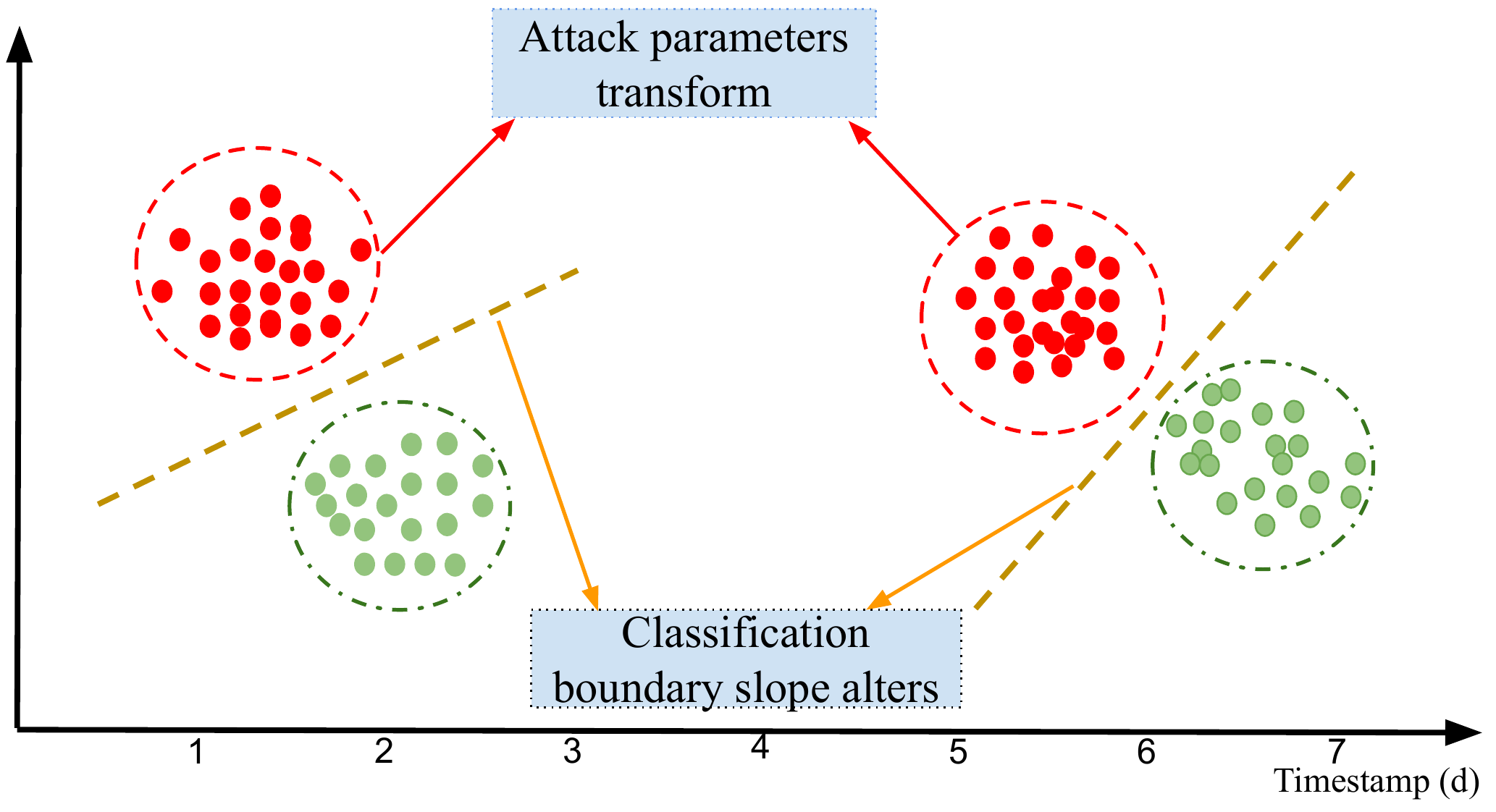}
    \caption{A scenario of mutative attack.}
    \label{fig:bs_flow}
    \end{minipage}%
\end{figure*}

\section{IoT Attack Detection Framework}
\label{sec:model}

In this section, we mainly describe the components of our attack detect framework in IoT networks, which is shown in Figure \ref{fig:framework}. First, we revisit the information theory for anomaly detection and propose to use the entropy-based metrics to distinguish the benign and malicious traffic in IoT networks. Then, we investigate the performance of distinct features that are utilized for IoT attack detection. Due to the emergence of learning-automation attacks, as shown in Figure \ref{fig:bs_flow}, the classification boundary between normal and malicious traffic may move when the attacker changes its attack parameters, such as attack rate and type. Therefore, we introduce a reinforcement learning model to our attack detection framework to continuously update the classification boundary, thereby well adapting to the new IoT attacks.

\subsection{Information Metrics for IoT Attack Detection}
Information metrics are adopted widely to discover the anomalies in intrusion detection fields \cite{hou2018latent, xie2016distributed }. In information theory, entropy measures the randomness of an information variable.  The value of entropy tends to be larger when the information variable shows to be more random. On the contrary, when the amount of uncertainty for the information variable is small, the entropy will have a relatively low value. The low entropy value demonstrates the concentration of distribution for information variables, which could reveal that there may exist an anomaly in the current system.

In this work, we attempt to apply the entropy-based approach to IoT anomaly detection. Shannon and Reyi entropy are the two typical metrics adopted in the information theory to measure the dispersion and concentration for the feature variable. We consider a discrete probability distribution:
\begin{equation}
P(X) = \{ p(x_{1}), p(x_{2}),...,p(x_{n}) \} \, \sum_{i=1}^{n}p(x_i) = 1
\end{equation}

The variable $X$ could be a feature that take the discrete values of $\{x_{1}, x_{2},...,x_{n}\}$. $p(x_{i})$ represents the probability that X takes value $x_i$. Then the Renyi entropy of order $\alpha$ is defined as:
\begin{equation} 
H_{\alpha}( X ) =  \frac{1}{1- \alpha }  log_{a}\big( \sum_{i=1}^{n} p(x_i)^{ \alpha } \big)
\end{equation}

When $\alpha \xrightarrow{} 1$, $H_{\alpha}(X)$ converges to Shannon entropy:
\begin{multline}\label{eq:entropy}
H_{s}(X) =  \lim_{ \alpha  \rightarrow 1} \big( \frac{1}{1-\alpha} log_{a}( \sum_{i=1}^{n} p(x_{i})^{\alpha} )\big) 
\\ = \sum_{i=1}^{n} p(x_{i}) log_{a}( \frac{1}{p(x_{i})} )
\end{multline}

The Shannon entropy of $X$ can also be regarded as the expected value of $ log_{a}( \frac{1}{p(x_{i})})$ with the probability mass function $p(x_i)$. The value of the entropy $H_{s}(X)$ reflects the randomness of the probability distribution and reveals that the feature has a more dispersed or concentrated property of probability distribution. 

In addition to measure the randomness of feature variable, the extent of changes between assumed and observed distribution could also be measured for discovering some anomalies. Let us use two discrete probability distribution $U$ and $V$ for the feature variable $X$. The information divergence between distribution of $U$ and $V$ of order $\alpha$ could be denoted as:
\begin{equation}
D_{\alpha}(U||V) =  \frac{1}{1 - \alpha}log_{a} \big( \sum_{i=1}^{n}  \frac{u({x_i})^{ \alpha }}{v({x_i})^{1- \alpha }}  \big)
\end{equation}
When $\alpha \xrightarrow{} 1$, $D_{\alpha}(U||V)$ converges to Kullback-Leibler divergence:
\begin{multline} \label{eq:kl}
D_{KL}(U||V) =  \lim_{ \alpha  \rightarrow 1}   \frac{1}{1 - \alpha}log_{a} \big( \sum_{i=1}^{n}  \frac{u({x_i})^{ \alpha }}{v({x_i})^{1- \alpha }}  \big) 
\\ =  \sum_{i=1}^{n} u(x_i)log_{a} \frac{u(x_i)}{v(x_i)} 
\end{multline}
In order to measure the randomness of the probability distribution only, which is independent of the number of distinct values of the feature variable, we need to normalize the entropy by dividing it with $log_{a}(N_0)$,
where $N_0$ is the number of distinct values for information variable. Then the entropy will be re-defined as follows:
\begin{equation}
    H_{ns}(X) = \frac{\sum_{i=1}^{n} p(x_{i}) log_{a}( \frac{1}{p(x_{i})} )}{log_{a}(N_0)} 
\end{equation}
Then the values of the entropy will fall in the range of $(0,1)$.

The Shannon entropy and Kullback-Liebler relative entropy could be both utilized for the attack detection in IoT networks. The information metric can overcome the limitations of other intrusion detection methods and can well adapt to the IoT attack detection scenario. We analyze the advantages of employing information metrics to IoT attack detection as follows:

\begin{itemize}
    \item The typical IoT attacks could be detected by analyzing the networking and application layer traffic. IoT gateway is the interface that bridges the IoT devices and IoT networks, which can aggregate a considerable amount of traffic. The entropy-based approach is more suitable for flow-based attack detection.
    
    \item Nowadays, there is a trend that most computations are moving to the edge in order to reduce the latency of services significantly. Compared to the massive computation power in the cloud, the edge usually has relatively low computation power. The entropy-based attack detection approach requires a lower computation cost and can be deployed in the edge and even on the IoT gateway.
    
    \item As more attacks have characteristics of covertness, the intrusion detection approach is required to have the capabilities of identifying the low-rate attack types. The entropy-based approach is more sensitive and can well distinguish the difference between legitimate traffic and malicious traffic using a minimum number of attributes.
    
    \item Due to the lower computation cost, the entropy-based approach can detect the attack in a real-time fashion. Furthermore, multiple instances can be running simultaneously due to its excellent scalability, which can be leveraged to monitor different types of IoT attacks. Such an approach also shows the capability of integration into our advanced reinforcement learning-based attack detection framework.
\end{itemize}
The above advantages will be further assessed and demonstrated in the evaluation section. Next, we start to analyze the features in IoT networks that could be used for the attack detection via entropy-based anomaly discovery.

\subsection{Feature Extraction for IoT Attack Detection} \label{sec:entropy_features}
We have introduced the entropy as the method to measure the feature variable and discover the potential IoT attacks. Let us take a \textit{ARP spoofing} as an example to explain how the entropy works to detect the attack. Then, we try to explore all the features that could unveil a variety of attacks.

ARP spoofing is a typical attack occurring in the IoT networks. The attacker sends falsified ARP (Address Resolution Protocol) messages over a local IoT network resulting in the link of the attacker's MAC address with the IP address of a legitimate IoT device on the network. Once the attacker's MAC address is bound to an IP address of an authenticated IoT device, the attacker will begin receiving any data that is intended for that IoT device, which can enable the attacker to collect the users' private data and also intercept, modify or even stop data-in-transit. Suppose we use the feature variable $X$ to represent the number of packets transmitted with different protocols during a time window t. For each protocol $x_i$, we can calculate the probability:\\
\begin{equation}
p(x_i) = \frac{\text{\# pkts transmitted with protocol } x_i}{\text{Total number of pkts transmitted}}
\end{equation}
The normalized factor is $log_{a}(N_0)$, where $N_0$ is the number of active networking protocols observed during the measured time window. In the window, due to the increased number of ARP packets transmitted caused by ARP spoofing, the probability distribution may show the characteristic of concentration resulting in the decrease of the entropy value. 

If the value of entropy is below the preset threshold $\theta^{*}$:
\begin{equation}\label{eqn:ent-thresh}
     H_{ns}(X) = \frac{\sum_{i=1}^{n} p(x_{i}) log_{a}( \frac{1}{p(x_{i})} )}{log_{a}(N_0)} < \theta^{*}
\end{equation}
That may indicate the presence of an anomaly and alert the defender. Furthermore, ARP spoofing attacks are often used to facilitate other IoT attacks such as Denial-of-Service (DoS) attacks. DoS attacks can utilize ARP spoofing to link the set of IP addresses from multiple IoT devices with a single target's MAC address. As a result, all the traffic that is intended for many IoT devices will be redirected to the target's MAC address, overloading the target with traffic. Such attacks are also similar to the TCP/UDP flooding, which immediately flood the requests to the target IoT devices and disable their services. To detect such kinds of attacks, the IoT gateway is the ideal place to aggregate the traffic and conduct the analysis.

In this work, we consider to detect a variety of attacks in IoT networks, including direct attack and reflection attack. The direct attack could be further categorized into network layer attack and application layer attack. Thus, we introduce more features to improve the capability of detection for various attacks.

For a IoT device in the network, it could work as a source address and send packets to multiple destination addresses. Also, it can be regarded as a destination and receive packets from other hosts with different source addresses. For a unique IoT device, we can calculate the probability that target device is the destination:
\begin{equation}
p(x_i) = \frac{\text{\# pkts transmitted from source address } x_i}{\text{Total number of pkts received}}
\end{equation}
And also calculate the probability that target device is the source:
\begin{equation}
p(x_i) = \frac{\text{\# pkts transmitted to destination address } x_i}{\text{Total number of pkts sent}}
\end{equation}

On the other hand, the packets transmitted from different ports can reveal some anomaly information used for attack detection. For the TCP/UDP flooding attack, some traffic flows are concentrated on a specific set of ports, resulting in the following entropy value of port feature decreasing to a level that causes alarm:
\begin{equation}
p(x_i) = \frac{\text{\# pkts transmitted with port } x_i}{\text{Total number of pkts transmitted}}
\end{equation}

However, for the PortScan attack, the attacker will send probe packets to random destination ports instead of a couple of ports, making the probability distribution more random and, therefore, increasing the entropy value. Once the entropy value exceeds a threshold $H_{ns}(X) > \theta^{*}$, it will also imply a potential attack. However, for simplification, we do not describe this case in our model.

Based on Formula \ref{eqn:ent-thresh}, the threshold $\theta$ is used to determine whether there has appeared an IoT attack. In most work, the value is pre-defined based on expert experience or historical observation. However, the pre-defined threshold will be immediately invalid if the attacker changes their attack vector or switch its attack approach. For instance, the attacker may transform its high-rate attack to a low-rate attack, which may increase the miss detection rate if the defender still uses its preset threshold.

Therefore, dynamically optimizing the threshold based on the detection's feedback is a better approach to increase the intrusion detection rate significantly. Furthermore, as the attackers evolve to be more intelligent, the fixed threshold cannot catch up with the dynamic changes of the attacks. Next, we will introduce reinforcement learning to our attack detection model to adapt to the new trend of attack detection.

\subsection{Augmented Attack Detection via Reinforcement Learning}
%two papers 
%1. Spoofing detection with reinforcement learning in wireless networks
%2. Detecting stealthy botnets in a resource-constrained environment using reinforcement learning.
\textit{\textbf{1) Why do we need Reinforcement Learning ?}}

\begin{figure}
    \centering
    \includegraphics[width=0.65\linewidth, height=35mm]{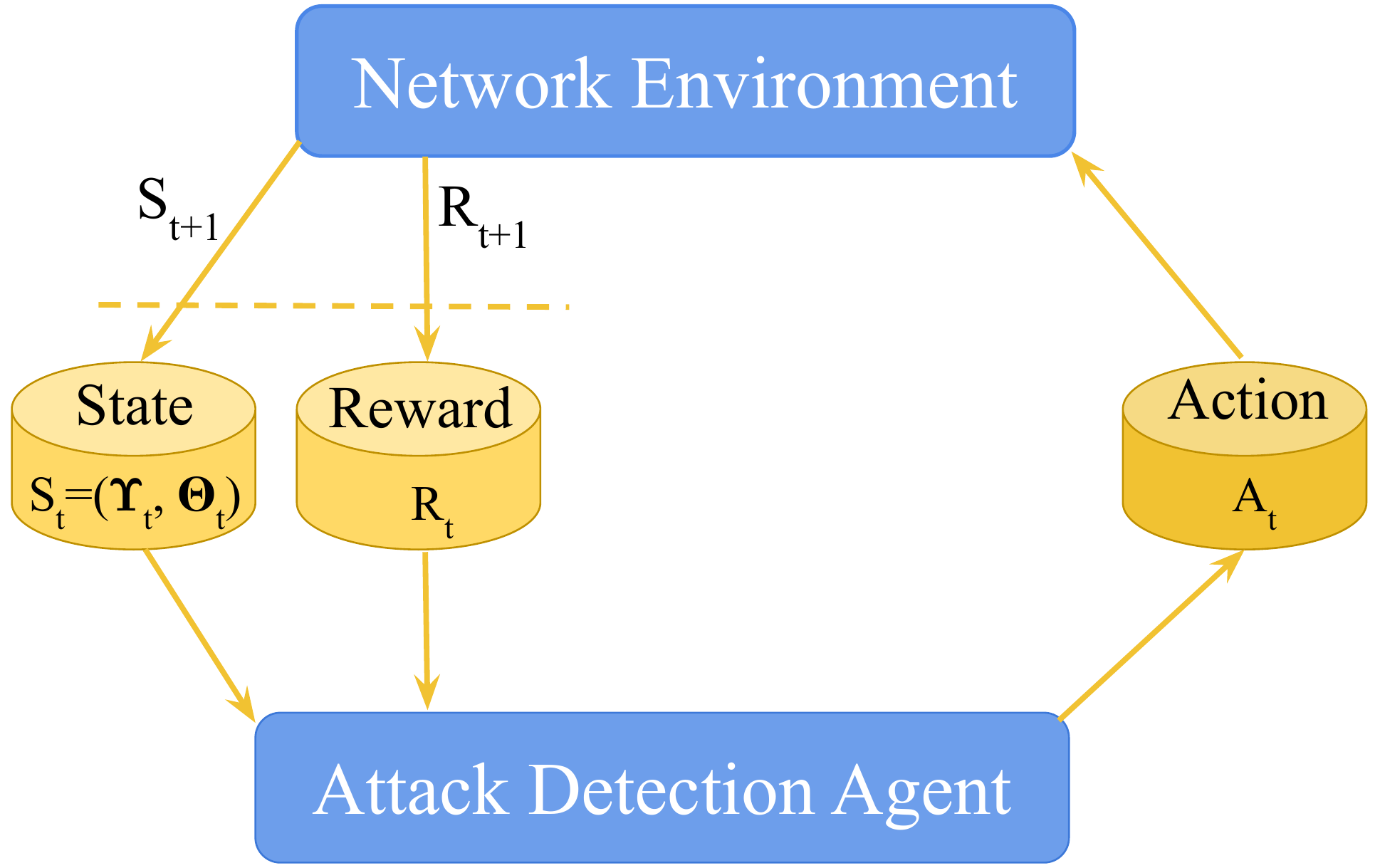}
    \caption{Iterative reinforcement learning process}
    \label{fig:rl_model}
\end{figure}

\textit{Reinforcement Learning(RL)} \cite{sutton2018reinforcement} is an algorithmic method for solving sequential decision-making problems wherein an agent (or decision-maker) interacts with the environment via iterative processes to learn \textit{how} to respond under different conditions, which is exhibited in Figure \ref{fig:rl_model}. Formally, the agent seeks to discover a policy that maps the system state to an optimal solution. Our goal is to train the agent to learn a policy that maximizes the total number of IoT attacks detected over time, thereby greatly increasing the system utility.

For IoT attacks, to persist in the IoT networks and obtain more attack profits, the attackers have to own the properties of \textit{stealth} and \textit{resilience}. To maintain the sneak of attacks, the attacker may leverage various methods to obfuscate the observable behaviors of attacks and evade detection. For instance, the attackers can periodically change the spatiotemporal properties of the traffic generated by attacks. With the development of AI and ML, the attackers have a trend to be more intelligent. Once the attacker finds that the specific attack is detected, the attackers could immediately and automatically adjust their attack strategy and continue to pursue attacks against IoT networks, leading to the possession of good resilience. The AI technique can even assist the attacker to automatically develop new attacks, which require the defender to have an advanced attack detection capability.

However, a defender typically has constraints on the number of available resources that can be used for attack detection, such as the limited computation power and human intervention. The defender's objective is to minimize the attack gain of the attacker using a limited number of resources, while the goal of the attackers is to obtain the greatest attack benefits by using fewer resources. Such a scenario can be modeled as a \textit{Markov Decision Processes}, which could be solved using reinforcement learning technique.

\vspace{2mm}
\textit{\textbf{2) Problem Statement}}

Suppose the attackers can launch a variety of attacks denoted by attack vector $\mathnormal{A}=(a_{1},a_{2},a_{3}, ..., a_{n})$. The thresholds used to decide the corresponding attacks in time slot $t$ can be represented by $\mathnormal{T^{t}}=(\theta_{a_{1}}^{t},\theta_{a_{2}}^{t},\theta_{a_{3}}^{t}, ..., \theta_{a_{n}}^{t})$. The elements in the attack vector are being observed simultaneously and continuously. Specifically, in time $t$, if the observed entropy $H_{a_{i}}^{t}$ for attack $a_{i}$ about a feature exceeds the current threshold $\theta_{a_i}^{t}$. The IoT attack detector sends an alert to the administrator. The detection results can be positive (identify the malicious traffic as an attack) or negative (identify the malicious traffic as not an attack). The detection results for each subject may or may not match the subject's actual status, which could be summarized as the following four cases:
\begin{itemize}
    \item True Positive(TP): Malicious traffic identified as an attack
    \item False Positive(FP): Benign traffic falsely identified as an attack
    \item True Negative(TN): Benign traffic identified as normal
    \item False Negative(FN): Malicious traffic falsely identified as normal
\end{itemize}
For the above different cases, there will be distinct utilities triggered by the environment. If the IoT attack agent reports a potential attack and hands it to the upper layer for further examination, the cost will be denoted as $-C_{0}$. Finally, if the reported alarm is identified as an attack, the detection system will receive the profits denoted as $P_{0}$. Otherwise, if the reported attack is identified as a false alert, the system will get a penalty $C_{1}$. If the real attack is not identified and let it pass through, the system will also get a penalty $C_{2}$. The last case is that the legitimate traffic is also detected as normal. The corresponding profit that can be obtained is denoted as $P_{1}$. The ultimate \textbf{system utilities} for different cases are shown in Table I.

Now we calculate the accumulated occurrences for each case during a time period $T$ comprising of $n$ time slots $(t_{1}, t_{2}, {t_3}, ..., t_{n})$. We use $N_{11}^{T}$ to represent the number of authenticated true attacks. The number of false alarm is denoted as $N_{12}^{T}$. The number of real attacks but missed is $N_{21}^{T}$. $N_{22}^{T}$ is used to denote the number of authenticated benign traffic flows. Then, in time period $T$, the complete reward obtained from the environment is:
\begin{equation}
R^{T} = (P_{0}-C_{0})*N_{11}^{T} - (C_0 + C_1)*N_{12}^{T} - C_2*N_{21}^{T} + P_1*N_{22}^{T}
\end{equation}
where
\begin{equation}
N_{11}^{T}+N_{12}^{T}+N_{21}^{T}+N_{22}^{T} = n    
\end{equation}
Then, the hit rate $\Upsilon_{T}$ and false alarm rate $\Theta_{T}$ could be computed as follows:

\begin{table}[]
\label{table:eb}
\caption{Environment Feedback}
\centering
\begin{tabular}{|l|l|l|l|}
\hline
True Positive: & $P_{0}-C_{0}$ & False Positive: & $-C_{0}-C_{1}$ \\ \hline
False Negative: & $-C_{2}$ & True Negative: & $P_{1}$ \\ \hline
\end{tabular}
\end{table}

\begin{equation}
\Upsilon_{T} = \frac{TP}{TP+FN} = \frac{N_{11}^{T}}{N_{11}^{T} + N_{21}^{T}}
\end{equation}
\begin{equation}
\Theta_{T} = \frac{FP}{FP+TN} = \frac{N_{12}^{T}}{N_{12}^{T} + N_{22}^{T}}
\end{equation}
The system state in time period $T$ could be represented as:
\begin{equation}
S_{T} = (\Upsilon_T, \Theta_{T})
\end{equation}
In this work, we model the attack detection in IoT networks as \textit{Markov Decision Processes}. The goal is to maximize the utility of IoT attack detection system $R_T$ via optimizing the threshold $\theta_{a_i}$ for detecting the specific attack type $a_i$.
\begin{equation}
\theta_{a_{i}}^{*}= \underset{\theta_{a_i} \geq 0}{\arg\max R_{T}}
\end{equation}
In order to maximize the system utility, the feature threshold needs to be appropriately chosen. We desire the detection system not to miss the detection of some attacks. On the other hand, we also do not want the system to report many false alarms, which may require massive human interventions. We need to stimulate the system to carefully and accurately conduct attack detection with corresponding rewards, thereby significantly reducing human interventions. Therefore, we propose a reinforcement learning-enabled attack detection approach to guide the defender's sequential decision-making process over time. 

\vspace{2mm}
\textit{\textbf{3) Reinforcement Learning based Attack Detection Model}}

To tackle this problem, we propose to formulate the problem as a \textit{Markov game} $G$ using reinforcement learning, which is defined by a tuple $G =\mathcal{(S,A,P,R)}$, where $\mathcal{S,A,P,R}$ are the sets of states, joint action space, transition probability functions, and reward functions respectively \cite{8714026}.

The detailed definitions are given as follows:

\begin{itemize}
    \item \textbf{State:} $\mathcal{S} \triangleq \{s^1,s^2,..., s^{\mathbb{A}}\}$ is the state space of the IoT attack detection agent. The system state $s_{T}$ in time period $T$ included in $\mathcal{S}$ is denoted as $s_{T}^i = (\Upsilon_T, \Theta_{T})$, which represents the hit rate and false alarm rate with the threshold $\theta_{a_i}$ in time period $T$. In order to compress the state space, we chunk the continuous values of $(\Upsilon, \Theta)$ and convert it to the discrete value pairs: $(\Upsilon_T, \Theta_T) \in \{ (\Upsilon^{i}, \Theta^{i}), i = (1,...,\mathbb{A})\}$. $\mathbb{A}$ is the size of the state space.
    %\vspace{2mm}
    \item \textbf{Actions:} 
    $\mathcal{A} \triangleq \{ a^1, a^2,..., a^{\mathbb{B}} \} $ is the action space of the IoT attack detection agent. Here, we split the threshold range for attack type $\theta_{a_i}$ into $\mathbb{B}$ discrete values. The action $\mathbb{A}^{j}_{T}(j \in \{1,...,\mathbb{B}\})$ specifies the selected threshold for the attack detection in time period $T$.
    %\vspace{2mm}
    \item \textbf{Reward function:} $\mathcal{R} \triangleq \mathcal{S} \times \mathcal{A} \rightarrow \mathbb{R}$ are the reward functions representing the utility of IoT attack detection agent obtained after the specific action is executed in a certain state. The agent attempts to choose the best action (namely threshold) based on its temporal state to maximize its expected discounted reward: $ \sum_{d=0}^{\infty} \gamma^{d} r_{t+d}$. The reward $r_{t}$ is defined as the utility when the agent executes the action $a_t$ in state $s_t$.
    %\vspace{2mm}
    \item \textbf{State transition probability:} $\mathcal{P} \triangleq \mathcal{S} \times \mathcal{A} \rightarrow [0,1]$ is the transition probability of the system. $p(s_{t+1}|s_{t}, a_{t})$ gives the probability of transiting to $s_{t+1}$ given a action $a_{t}$ is taken in the current state $s_{t}$.
    %\textcolor{red}{add more details to explain in current attack context.}
\end{itemize}

\begin{algorithm}[]
\SetAlgoLined
\textbf{Input:} learning rate $\alpha$, discount factor $\gamma$, $\epsilon$-greedy, number of time slots $\mathbb{N}$ in a episode, initialize all the $Q$ table entries $Q(s,a)$ to zero, initialized state $s$.\\
%\textbf{Repeat} (for each episode)\\
\Repeat{Attack detection is terminated}{
    Start a new episode.\\
    Observe current system state $s =(\Upsilon, \Theta)\in \mathcal{S}$.\\
    Select a threshold $\theta^{*} = \underset{a\in \mathcal{A}}{\arg\max} Q(s,a)$.\\
    \For{$i\gets1$ \KwTo $\mathbb{N}$}{
        Compute the entropy value via equation \ref{eq:entropy}.\\
        \eIf{$H_{ns}(X) < \theta^{*} $}{
        %instructions1\;
            Let the traffic pass through.
        }{
        Report the potential IoT attack alarm .\\
        Further screen the alarm by \textit{Defender}.\\
        }
    }
    Obtain the reward $r$ and transform to a new state $s'$.\\
    Update the table entry for $Q_{t+1}(s,a)$ to $Q_{t}(s,a) + \alpha \big[ r_{t}(s,a) + \gamma \underset{a'}{max}Q_{t}(s',a') - Q_{t}(s,a) \big]$.\\
    Replace $s$ with $s'$.
}
 \caption{IoT Attack Detection with Q-learning}
\end{algorithm}

In our model, the attack detection agent starts at some initial state $s^{i} \in \mathcal{S}$. After observing the current state, the agent selects the action $a^{j} \in \mathcal{A}$ and the agent will receive the corresponding rewards together with the new observation. At the same time, the system will transit to a new state $s' \in \mathcal{S}$ with probability $p(s'|s,a)$. The procedure is repeated at the new state and continues for a finite or infinite number of stages. The agent tries to find its optimal policies to maximize the expected long-term average rewards. In the context of IoT attack detection, the attack detection agent attempts to continuously optimize the threshold used for attack detection in consecutive time periods and strives to increase the detection rate and decrease the false alarm rate, thereby maximizing the long-term average system utility. 

\vspace{2mm}
\textit{\textbf{4) IoT Attack Detection via Q-Learning}}

In a reinforcement learning process, an IoT attack detection agent can learn its optimal policy (namely threshold) through interaction with its environment. In particular, the agent first observes its current state, and then takes action, and receives its immediate reward together with its new state. The current state is the attack detection rate and false alarm rate under the present detection threshold. The selection of a new threshold is executed as action and will lead to the updated system reward with the new detection and false alarm rate which is regarded as a new state. The observed information, i.e., the immediate reward and new state from the environment feedback, is also used to adjust the agent’s policy, and this process will be repeated until the agent’s policy approaches to the optimal policy. In reinforcement learning, Q-learning \cite{van2016deep} is the most effective method and widely used in reinforcement learning. In the following, we will discuss how to use the \textit{Q-learning} algorithm to implement IoT attack detection.
\begin{table*}[]
\centering
\caption{Description of IoT Attack Dataset.}
\label{tab:my-table}
\begin{tabular}{|c|c|c|c|c|}
\hline
IoT Devices & \multicolumn{4}{c|}{\begin{tabular}[c]{@{}c@{}}WeMo motion,  WeMo Switch, Samsung smartcam, TP-Link smart plug, Netatmo Camera\\ Chromecast Ultra, Amazon Echo, Phillips Hue bulb, iHome smart plug, LiFX bulb\end{tabular}} \\ \hline
\multirow{2}{*}{Attack Types} & \multicolumn{2}{c|}{Direct Attack} & \multicolumn{2}{c|}{Reflection Attack} \\ \cline{2-5} 
 & \multicolumn{2}{c|}{\begin{tabular}[c]{@{}c@{}}ARP spoofing, TCP SYN Flooding, \\ UDP Flooding, and Ping of Death\end{tabular}} & \multicolumn{2}{c|}{SNMP, SSDP, TCP SYN, and Smurf} \\ \hline
\multirow{2}{*}{Attack Parameters} & Number of Attacks & 200 times & Every Attack Time & 10min \\ \cline{2-5} 
 & Attack Rates & 1, 10, 100 pkts/second & Time Span & 16 days \\ \hline
\end{tabular}
\end{table*}
In Q-Learning Algorithm, we aim to find an optimal entropy threshold $\theta^*$ for the attack detection agent to maximize the system utility. Accordingly, we first define value function $\mathcal{V}^{\theta}: \mathcal{S} \rightarrow \mathbb{R}$ that represents the expected value obtained by the threshold $\theta$ from each state $s \in \mathcal{S}$.
%\vspace{-3mm}
\begin{multline}
\mathcal{V}^{\theta}(s)=\mathbb{E}_{\theta} \Big[ \sum_{t=0}^{\infty }  \gamma r_{t}(s_t, a_t)|s_{0} = s \Big] \\
= \mathbb{E}_{\theta} \Big[ r_{t}(s_{t},a_{t}) + \gamma \mathcal{V}^{\theta}(s_{t+1} \big| s_{0}=s \Big]
\end{multline}
If we denote $Q^{*}(s,a) \triangleq r_{t}(s_t,a_t) + \gamma \mathbb{E}_{\theta}[\mathcal{V}^{\theta}(s_{t+1})]$ as the optimal Q-function for all state-action pairs, then the optimal value function can be written by $\mathcal{V}^{*}(s) = \underset{a}{max} \{ Q^{*}(s,a)\}$. Now the problem is reduced to find optimal values of Q-function, i.e., $Q^{*}(s,a)$, for all state-action pairs, and this can be done through iterative processes. In particular, the Q-function is updated according to the following rule:
\begin{multline}\label{q-iterate}
Q_{t+1}(s,a) = Q_{t}(s,a) + \\
\alpha_{t} \Big[ r_t(s,a) + \gamma \underset{a'}{max} Q_{t}(s,a') - Q_{t}(s,a)\Big]
\end{multline}
In Formula \ref{q-iterate}, the learning rate $\alpha_{t}$ is used to determine the impact of new information to the existing Q-value. The learning rate can be chosen to be a constant, or it can be adjusted dynamically during the learning process. The algorithm yields the optimal policy indicating an action to be taken at each state such that $Q^{*}(s, a)$ is maximized for all states in the state space, i.e., $\theta^{*}(s) = \underset{a}{\arg\max} Q^{*}(s,a)$, which is displayed as a Q-Table. The detailed procedures are described in the Algorithm I. Our model continually updates the feature vectors for various attacks encountered such that it can recognize changing attack patterns. Our model also memorizes and retains the knowledge of previous attacks learned. Once the attacker changes to the attack pattern that appears before, our model can fast identify it without executing the reinforcement learning iterative process.

\begin{figure*}
\centering  
\subfigure[TP-Link smart plug]{\includegraphics[ width=0.245\linewidth]{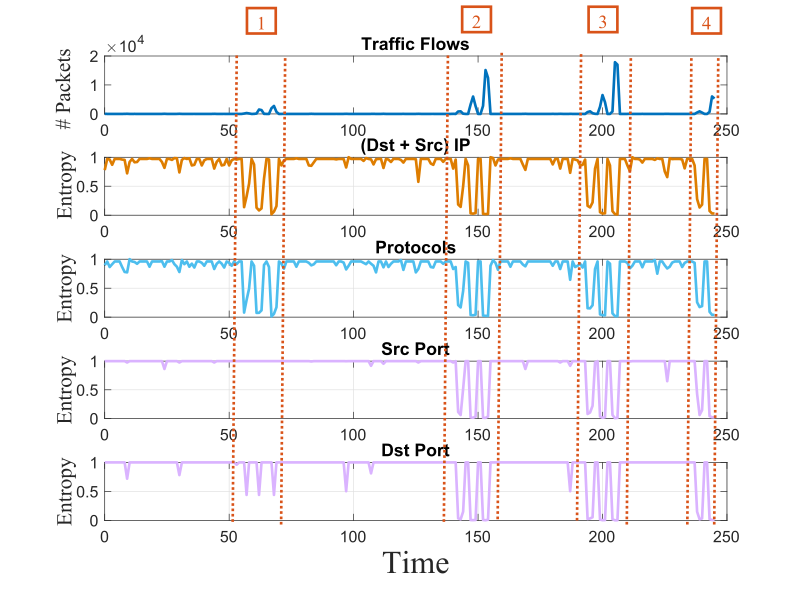}}
\subfigure[Samsung smartcam]{\includegraphics[ width=0.245\linewidth]{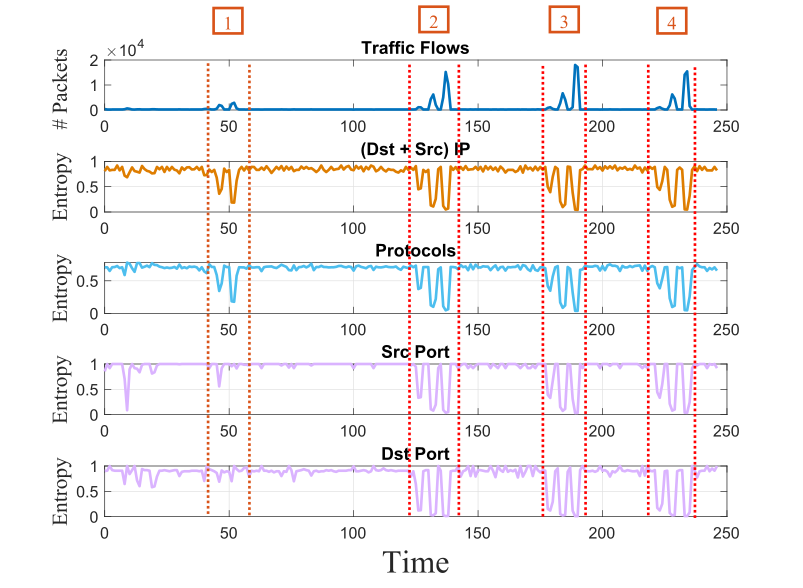}}
\subfigure[WeMo power switch]{\includegraphics[ width=0.245\linewidth]{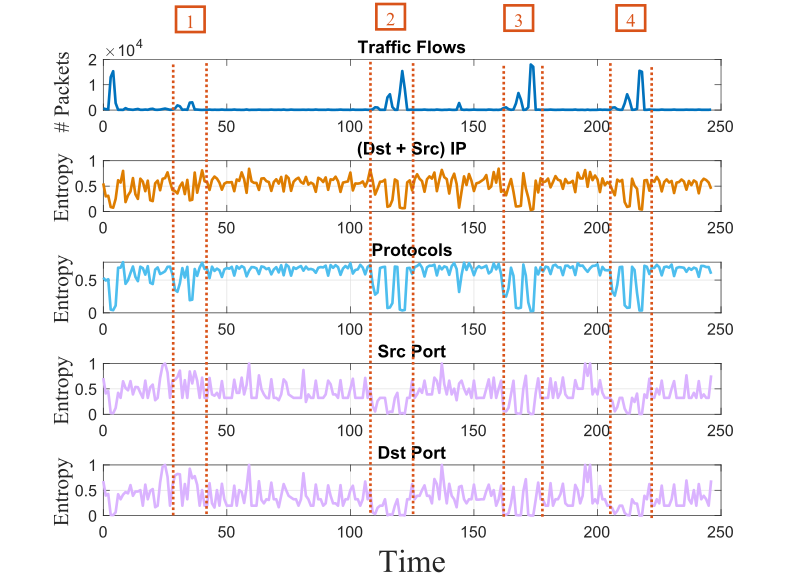}}
\subfigure[WeMo motion sensor]{\includegraphics[ width=0.245\linewidth]{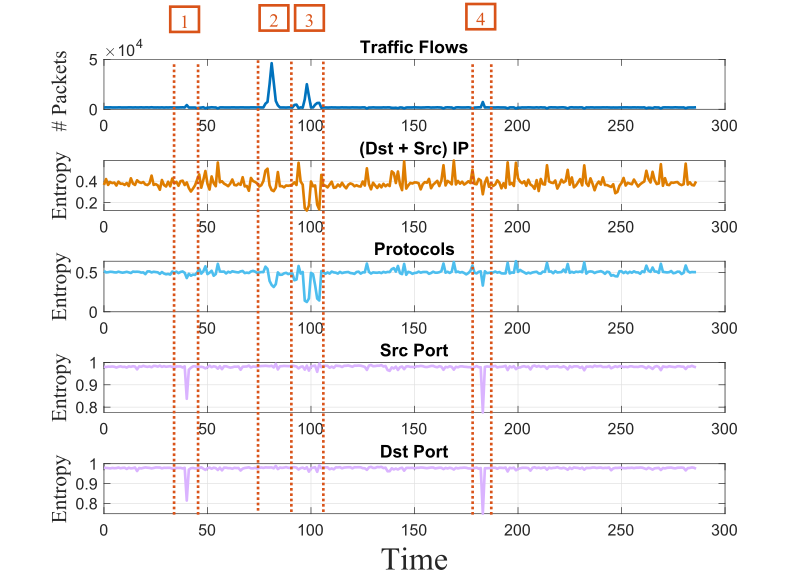}}
\caption{Time series of entropy data for four IoT devices with different features.}
\label{fig:entropy_analysis}
\end{figure*}

% \begin{figure*}
%     \centering
%     \begin{subfigure}[b]{0.245\textwidth}
%         \includegraphics[width=\textwidth]{Figures/tp_link.eps}
%         \caption{TP-Link smart plug}
%         \label{fig:tp_link}
%     \end{subfigure}
%     %
%     \begin{subfigure}[b]{0.245\textwidth}
%         \includegraphics[width=\textwidth]{Figures/samsungcamera.eps}
%         \caption{Samsung smartcam}
%         \label{fig:samsung_camera}
%     \end{subfigure}
%     %
%     \begin{subfigure}[b]{0.245\textwidth}
%         \includegraphics[width=\textwidth]{Figures/powerswitch.eps}
%         \caption{WeMo power switch}
%         \label{fig:power_switch}
%     \end{subfigure}
%     %
%     \begin{subfigure}[b]{0.245\textwidth}
%         \includegraphics[width=\textwidth]{Figures/motionsensor.eps}
%         \caption{WeMo motion sensor}
%         \label{fig:motion_sensor}
%     \end{subfigure}
%     \caption{Time series of entropy data for four IoT devices with different features.}
%     \label{fig:entropy_analysis}
% \end{figure*}

\section{Evaluation}
\label{sec:eval}
In order to verify the effectiveness of our IoT attack detection framework, we design and conduct extensive experiments to thoroughly evaluate our approach over an IoT attack dataset from a real IoT platform \cite{hamza2019detecting}. First, we use the real IoT attack traces to analyze the performance of our proposed entropy-based anomaly detection features and models. Then, we use a real scenario that the attackers alter their attack patterns among 200 attacks to avoid intrusion detection. We apply our reinforcement learning-based attack detection model to this scenario and demonstrate our technique is capable of learning and recognizing changes in attack patterns, allowing it to adapt to intelligent or dynamically changing attacks, unlike traditional static threshold models. We show the evaluation in detail as following.

\subsection{Data Set}
The IoT dataset \cite{hamza2019detecting} is generated by a IoT platform comprising a TP-Link gateway connecting with ten IoT devices, such as TP-Link smart plug, Samsung smart-cam, WeMo motion sensor, etc. The vulnerabilities of these devices are investigated and identified first, and then a variety of attacks are launched including direct attacks, such as ARP spoofing and TCP flooding, and reflection attacks, such as SNMP and Smurf. A total of 200 attacks are launched to different IoT devices and that last for 16 days.

% \begin{figure}
%     \centering
%     \begin{subfigure}[b]{0.24\textwidth}
%         \includegraphics[width=\textwidth]{Figures/protocol_dist.eps}
%         \caption{Protocols.}
%         \label{fig:protocol_distribution}
%     \end{subfigure}
%     %
%     \begin{subfigure}[b]{0.24\textwidth}
%         \includegraphics[width=\textwidth]{Figures/ip_distribution.eps}
%         \caption{IP addresses.}
%         \label{fig:ip_distribution}
%     \end{subfigure}
%     %
%     \caption{Distribution of transmitted packets from different protocols and IP addresses with and without IoT attacks.}
% \end{figure}

\begin{figure}
\centering  
\subfigure[Protocols]{\includegraphics[ width=0.24\textwidth]{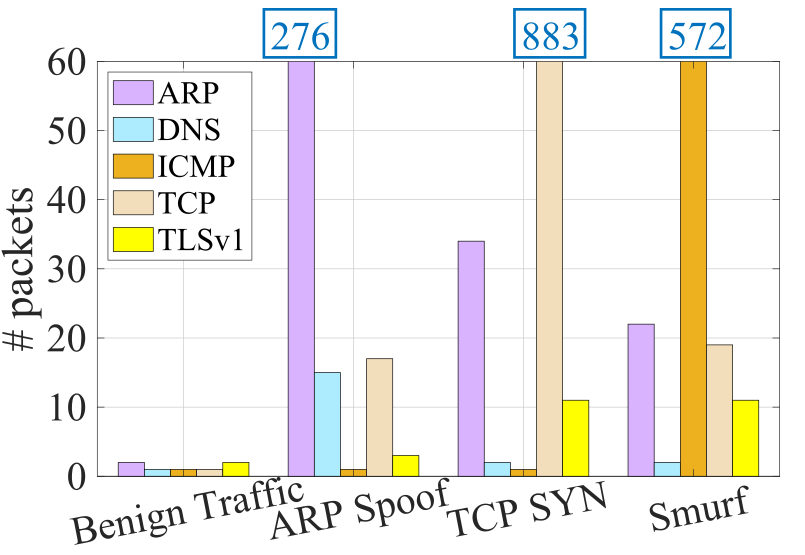}}
\subfigure[IP addresses]{\includegraphics[ width=0.24\textwidth]{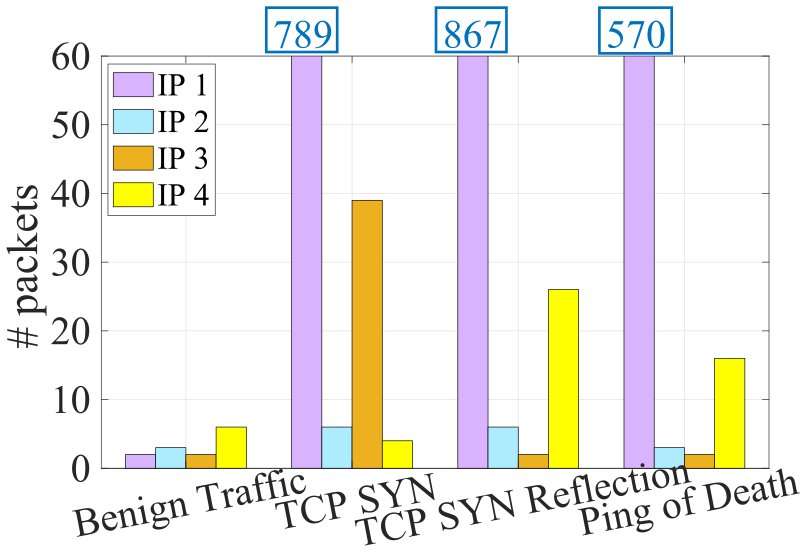}}
\caption{Distribution of transmitted packets from different protocols and IP addresses with and without IoT attacks.}
\label{fig:distribution_all}
\end{figure}

\begin{figure*}
    \begin{minipage}[c]{0.22\linewidth}
    \includegraphics[height=33mm]{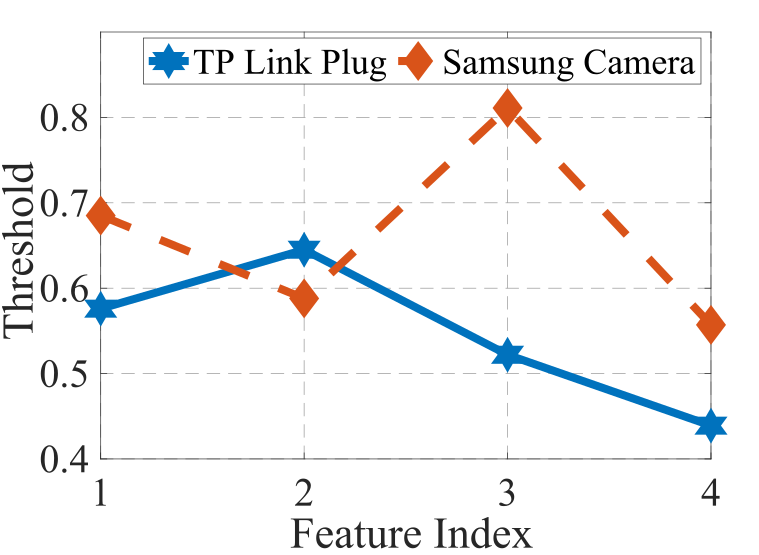}
    \caption{Snapshot of thresholds for attack detection.}
    \label{fig:snapshot_threshold}
    \end{minipage}
    \hfill
    \begin{minipage}[c]{0.48\linewidth}
    \includegraphics[height=33mm]{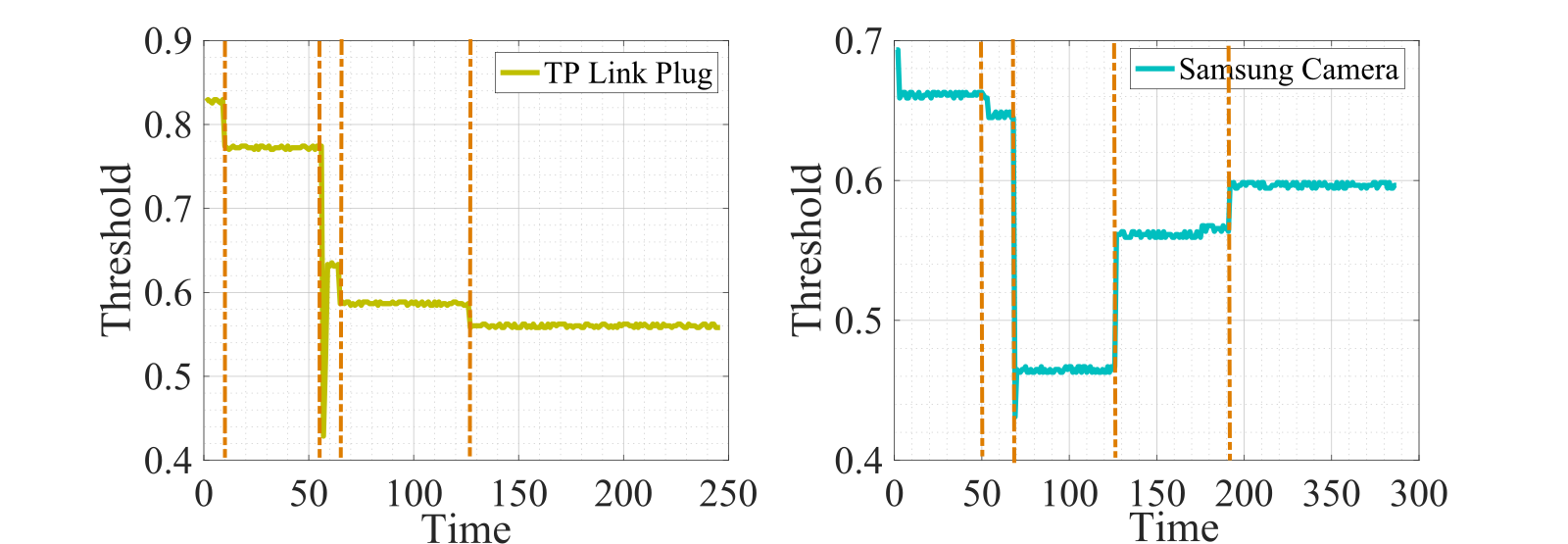}
    \caption{Reinforcement learning-based feature threshold optimization for attack detection.}
    \label{fig:threshold_sequences}
    \end{minipage}
    \hfill
    \begin{minipage}[c]{0.26\linewidth}
    \includegraphics[height=32mm]{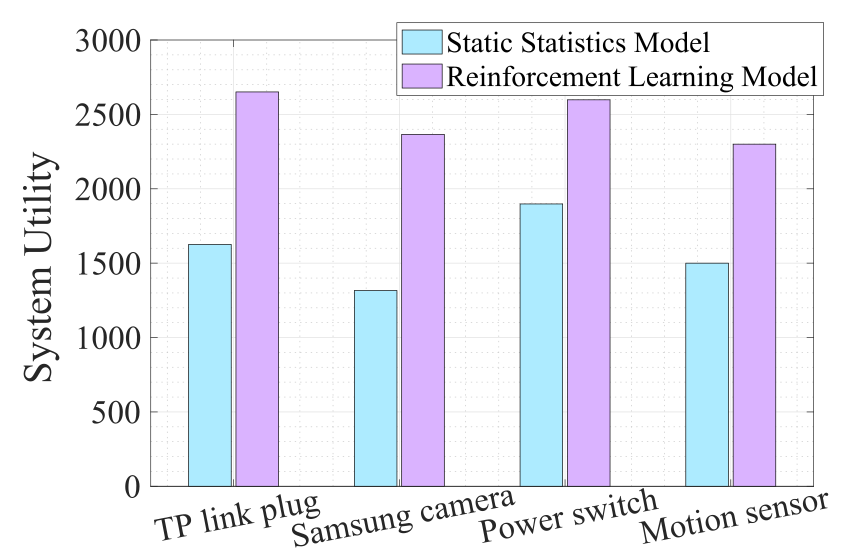}
    \caption{Comparison of system utility with and without RL.}
    \label{fig:utility}
    \end{minipage}%
\end{figure*}
The entire packet traces of benign and malicious traffic are stored in 30 \textit{pcaps} files and each \textit{pcaps} contains the traffic traces with a size of $2-3$ GB over a day. Also, the attacks are launched with three rates 1, 10, and 100 packet-per-second (pps), which can reveal the attack effect for both low-rate and high-rate. For each device, the attacks with different types and rates are launched. The detailed parameters about the dataset are described in Table II. The low-rate attacks confuse the distinction between benign and malicious traffic, and the transformation of attack type and rate also significantly boost the difficulty of attack detection. The properties of the dataset can thoroughly exam the performance of our IoT attack detection technique.

\subsection{Entropy-based feature analysis}
There exist some research works that apply entropy analysis to anomaly detection in cybersecurity. However, they use the traffic traces from traditional Internet mainly comprising of computers and servers to conduct analysis. No work considers the unique characteristics of IoT networks and uses the real IoT dataset for attack detection via using the entropy-based technique. In this work, we revisit the entropy-based anomaly detection approach and explore the effect when it is applied to IoT attack detection.

We evaluate the detection effect of various features we propose in Section \ref{sec:entropy_features}. The IoT attacks can be launched in different layers that may impact the traffic for distinct protocols. So the distribution of protocol-related packets between normal and malicious traffic is likely to be distinct. If the packets' distribution concentrates on a certain protocol, that may indicate the presence of an attack. Figure \ref{fig:distribution_all}(a) shows the distribution of packets for protocols \textit{(ARP, DNS, ICMP, TCP, and TLSv1)} with and without attacks. The attacks include \textit{ARP spoof, TCP SYN, and Smurf}. The data of the Figure is computed based on the trace set on 06/01/2019 for TP-Link smart plug. For the benign traffic, the distribution of packets about different protocols displays randomness resulting in the low entropy value. But the three attacks display the less randomness and more aggregation for some protocol, which will lead to the larger entropy value. Figure \ref{fig:distribution_all}(b) displays the distribution of packets under four distinct IP addresses with and without attacks. The attacks include \textit{TCP SYN, TCP SYN Reflection, and Ping of Death}. The figure reveals that the attacks lead to the explosion of packets in some IP address, which also results in much larger entropy value and manifests potential anomaly.

Moreover, we compute the entropy values of features with 5 min period using the traffic traces from TP-Link smart plug, Samsung smart-cam, WeMo power switch on 06/01/2019 and WeMo motion sensor on 06/02/2019, which is shown in Figure \ref{fig:entropy_analysis}. The four attacks, namely \textit{ARP Spoof (layer: L2D), TCP SYN (layer: L2D), TCP SYN Reflection (layer: W2D2W) and TCP SYN Reflection(layer: L2D2L)}, are launched to the first three devices. Another four attacks, namely \textit{UDP Flood (layer:L2D), SSDP (layer:D2W), SSDP (layer:L2D2L) and UDP Flood (layer:W2D)} are launched to the WeMo motion sensor.

Figure \ref{fig:entropy_analysis} labels all the launched attacks, and we have some findings from the figure. As shown, the entropy of IP and protocol are sensitive to most attacks and can achieve a good detection effect. However, for the UDP Flood attack, the difference of entropy values for benign and malicious traffic is not apparent, which is revealed in Figure \ref{fig:entropy_analysis}(d). Nevertheless, the entropy of port between benign and malicious shows a more substantial difference, which can be better utilized for detecting the attack. The benefit of entropy is that it can detect the attack even with the low-rate attack. For the ARP spoofing attack, we can find that the entropy of IP and protocol is still able to show the apparent fluctuation even with relatively lower transmission rates. For the WeMo power switch, the difference of entropy values between normal and malicious seems not to be much more apparent than other devices, but the threshold can still be appropriately chosen to distinguish the normal and anomaly traffic. Therefore, we combine features and build the feature vectors to detect the attacks robustly. Next, we evaluate our reinforcement learning-based detection technique that dynamically chooses the threshold for feature vectors to determine the attacks.

\subsection{Performance of reinforcement learning model}
Our reinforcement learning-based attack detection technique is to optimize the feature threshold to determine the attack continuously. The new trend that the attacker is evolving to be more intelligent requires the defender to develop the capability to adjust their defense strategy based on the variation of attacks. The traditional method relies on data statistics to determine a proper threshold at a given time. However, such method requires more human interventions to choose a new threshold once the attacker changes their attack patterns. Our approach can smartly and automatically detect attacks with less human supervision and significantly lower detection delay.

We design experiments to analyze the performance of our proposed RL-based IoT attack detection and compare it with the traditional attack detection with static threshold models. For the experiments, we set the values of parameters $P_0$, $P_1$, $C_0$, $C_1$, $C_2$ in Table I to be 14, 12, 0, 3, and 15. The parameters define the penalty and reward that our reinforcement learning model can obtain.  For the parameters about Q-Learning, the values of $\alpha$, $\gamma$ and $\epsilon$ are set to 0.1, 0.8 and 0.9 respectively. We stimulate the model to report the alarm accurately. If the model reports one alarm that is identified as an attack, the model can obtain the appropriate incentive. Otherwise, it receives the corresponding penalty.

We maintain a threshold vector for the same type of devices to store the threshold for each feature used for detecting the anomaly. Different types of IoT devices may have a distinct threshold vector due to their firmware, driver, and software heterogeneity. Figure \ref{fig:snapshot_threshold} shows a snapshot of the feature threshold at some point for TP-link plug and Samsung camera, respectively. The feature index denotes a part of the features we propose in Section \ref{sec:entropy_features}. For the same IoT device, the thresholds for each feature are also distinct. For static threshold models, the threshold vector is determined at the beginning and will not be changed unless human intervenes. Our learning agent continuously optimizes and updates the threshold vector for attack detection with much less human intervention. The continuously changing threshold for the feature IP address over a day is displayed in Figure \ref{fig:threshold_sequences}. Because we change the type of attacks and also adjust their attack rate, the corresponding boundary to distinguish benign and malicious traffic may be changed, leading to the variation of the threshold. The vertical dotted lines in the figure label the time point the attacker changes its attack type or attack rate. As shown, the reinforcement learning model can automatically adapt to the transformation of attacks and make adjustments for the threshold accordingly.

The utility of the IoT attack system is significantly improved because of the better detection and false alarm rate resulting from the continuous optimization of the threshold. We investigate the state-of-the-art work about IoT attack detection. We are the first to introduce the reinforcement learning technique to the detection model. Most research works like \cite{mirsky2018kitsune} use machine learning or deep learning to train an attack model and use it for future detection. We still regard them as the static statistics model because they do not continually update their model. We apply the static statistics model and our reinforcement learning model to detect the attacks for four devices. For each device, a certain number of attacks are launched with different attack patterns. Figure \ref{fig:utility} shows the utility comparison over a day for the four devices. The system utility can achieve an average 54.7\% increase across four devices when using the reinforcement learning technique compared to systems without it. With the change of the IoT attack pattern, the static threshold may gradually deviate from the optimal value resulting in bad detection rate and high false alarm rate, which will significantly reduce the system utility. Our reinforcement learning model is robust to the change of the IoT attack pattern and can dynamically update the feature threshold and significantly boost the attack detection rate to 98.5\%. Our IoT attack detection framework can also be widely applied for other intrusion detection fields and facilitate the current research in IoT security and reinforcement learning.
\section{Related Work}\label{sec:relatedwork}
\subsection{IoT security and privacy}
As the development and broadly deployment of IoT networks, more and more vulnerabilities involved in IoT platforms, applications, protocols, and hardware are being exposed to the attackers, which causes the critical security and privacy issues. \cite{celikiotguard} proposes a mechanism to protect the users from insecure devices states via monitoring the IoT behaviors and their corresponding platform apps. Fine-grained context identification \cite{jia2017contexlot} is introduced in the IoT platform to analyze the sensitive actions, thereby providing effective access control. %SmartAuth \cite{tian2017smartauth} uses Natural Language Processing (NLP) to extract the security information from an IoT app's description, code, and annotations, and then compare to the real actions that the app performs. 
For some vulnerabilities caused by the chain of multiple IoT components, the authors redefine the IoT system as multiple layers and present ForeSee \cite{9077392}, a cross-layer formal framework to
precisely capture potential attack paths in IoT systems. 

The communication protocols, such as ZigBee, Z-Wave, and BLE used in IoT networks, also bring out some security and privacy issues. For instance, the BLE protocol fails to hide the device's presence from the adversaries \cite{fawaz2016protecting}, which will lead to severe threats, such as user behavior tracking and inference of sensitive information. The authors introduce a new concept \textit{wireless context} \cite{8737451}, which represents the real IoT apps' workflows from the view of wireless communication. The potential system anomalies and hidden vulnerabilities can be detected by comparing the wireless context and the IoT context which is extracted from the IoT apps' descriptions via NLP. In the aspects of hardware and software, the fingerprinting technique \cite{gu2018bf} is proposed to secure the authentication and communication of IoT devices. 

\subsection{Machine learning in IoT networks}
To develop a secure, robust, and optimized solution, machine learning is introduced and is now widely adopted to solve issues in IoT networks \cite{benkhelifa2018critical}. \cite{chaabouni2019network,moustafa2018ensemble} 
use the machine learning technique to detect the attacks in IoT networks. IoT Sentinel \cite{miettinen2017iot} analyzes the networking traffic and uses the classification algorithms to identify the types of devices, enabling enforcement of rules for constraining the communications of vulnerable devices. A deep neural network-based framework is proposed in \cite{chatterjee2019rf} to implement the real-time nodes authentication in wireless networks. GCN (Graph Convolutional Networks) based anomaly detection model 
is proposed in \cite{9020760}
to detect anomalous behaviors of users, which can capture the graph relations of data.

Besides, due to the excellent learning characteristics, reinforcement learning is more suitable for some networking scenarios.\cite{chinchali2018cellular} proposes an RL-based scheduler that can dynamically adapt to traffic variation, and various reward functions set by network operators, to optimally schedule IoT traffic. \cite{ferdowsi2019deep} provides a security solution that uses deep reinforcement learning for signal authentication of IoT systems. \cite{xiao2016phy} proposes a spoofing detection mechanism in wireless networks that also leverages reinforcement learning to increase detection accuracy. \cite{lopez2020application} 
also proposes a deep reinforcement learning-based intrusion detection model for the traditional network. As we can see, there is no research work considering the characteristics of intrusion in modern IoT networks. We explore the attack characteristics in IoT networks and propose a dynamical intrusion detection model that can adapt to the transformation of intrusion strategies from attackers.

\section{Conclusion}\label{sec:conclusion}

In this paper, we explore the unique characteristics and trends of attacks in IoT networks. We argue that the traditional intrusion detection approaches in cybersecurity cannot adapt well to the IoT scenario. Due to the heterogeneity of platforms, protocols, software, and hardware, various vulnerabilities are exposed to attackers resulting in the increased complexity of anomaly detection. Meanwhile, the new emerging low-rate attacks obfuscate the boundary between benign and malicious attacks. Furthermore, the attacker is evolving to have the capabilities of altering the attack strategies and even developing a new attack based on the environment feedback, which requires the defender to have a quick response. Therefore, we propose an entropy-based attack detection framework integrating the reinforcement learning model, which is more suitable for the IoT context. We conduct extensive experiments over a real IoT attack dataset and demonstrate our IoT attack detection approach's efficiency.

\bibliographystyle{ieeetr}
\bibliography{references}
\end{document}